\documentstyle[preprint,revtex,eqsecnum]{aps}

\tightenlines
\begin{document}
\draft

\preprint{DOE/ER/40322-155, U. of MD PP \#92-193}

\begin{title}
Response of nucleons to external probes in hedgehog models: \\
I. Electromagnetic polarizabilities
\end{title}

\author{Wojciech Broniowski \cite{ifj} and Thomas D.  Cohen}
\begin{instit}
Department of Physics and Astronomy, University of Maryland \\
College Park, Maryland 20742-4111
\end{instit}

\begin{abstract}
Electromagnetic polarizabilities of the nucleon are
analyzed in a hedgehog model with quark and meson degrees of freedom.
Semiclassical methods are used (linear response theory,
quantization via cranking). It is found that
in hedgehog models (Skyrmion, chiral quark models,
Nambu--Jona-Lasinio model),
the average electric polarizability of the nucleon,
$\alpha_N$, is of the order $N_c$, and the
splitting of the neutron and proton electric(proper) polarizabilities,
$\delta\alpha = \alpha_n - \alpha_p$, is of
the order $1/N_c$. We present a
general argument why one expects  $\delta\alpha > 0$ in models with
a pionic cloud. Our model prediction for the sign and magnitude
of $\delta\alpha$ is in agreement with recent
measurements. The obtained value for $\alpha_N$,
however, is roughly a factor of three too large.
This is because of two problems with our particular model:
a too strong pion tail,
and the degeneracy of $N$ and $\Delta$
states in the large-$N_c$ limit. This
degeneracy also results in a very strong
$N_c$-dependence of the paramagnetic part of the
magnetic polarizability, $\beta$, which is of the order
$N_c^{3}$. We compare the large-$N_c$
results to the one-loop chiral
perturbation theory predictions, and show the importance
of $\Delta$-effects in pionic loops.
 We also investigate the role of non-minimal
substitution terms in the effective lagrangian
on the polarizabilities of the nucleon.
\end{abstract}

\pacs{PACS numbers: 12.38.Lg, 12.40.Aa, 14.20.Dh, 14.60.Fz }

\narrowtext
\section{Introduction}
\label{se:intro}

Polarizabilities are important fundamental properties of particles -- they
determine dynamical response of a bound system to external
perturbations, and provide valuable insight into internal strong-interaction
structure. Recent measurements of electromagnetic
polarizabilities of the proton \cite{Federspiel,Zieger} and neutron
\cite{Schmiedmayer} renewed theoretical interest in these quantities
\cite{olderexp}.
Many calculations of the electric, $\alpha$, and magnetic, $\beta$,
polarizabilities of the nucleon can be found in the literature in models
ranging from various quark models and bags
\cite{DMP,Hecking,Krivine,Schafer,Capstick} cloudy
bags \cite{Weiner85} to Skyrmions \cite{Nyman84,Chemtob87,Scoccola90}.
A quenched lattice calculation for $\alpha$ has also been done
\cite{Fiebig:lattice}.
Extensive reviews of the subject are available
\cite{Petrunkin81,Friar:rev}.

In this paper we analyze  the nucleon
polarizabilities in hedgehog models, using
the framework of the linear response theory.
Our analysis differs significantly from previous works by allowing
the soliton to deform (
dispersive effects). Our semiclassical methods are
described in detail in the following paper \cite{BC92partII},
referred to as (II).
The reader who is not familiar with the basics of hedgehog
models or semiclassical methods used in their treatment, is
urged to read (II) before this paper.

In recent years numerous hedgehog models, such as the Skyrme
model \cite{Skyrme6162,ANW83,Adkins84,skyrmion:rev}, chiral quark models
\cite{BirBan8485,KRS84,KR84,EisKal,BBC:rev,MCB:rev},
hybrid bag models \cite{hybrid:rev},
chiral models with confinement
\cite{BBC:rev,MKB:conf,BrLj,Duck}, or
the Nambu--Jona-Lasinio model \cite{NJL} in the solitonic treatment
\cite{Dyakonov86,DPPob88,MAGGM,Reihardt88,Meissner89,DPPr89,Alkofer},
were quite successful
in describing the phenomenology of nucleon structure. The basis
for these hedgehogs models is the large-$N_c$
limit \cite{thoft:nc,witten:nc} of QCD. Thus, $1/N_c$ should
consistently be treated as an expansion parameter in this approach.
Previous analyses of polarizabilities in hedgehog models
\cite{Nyman84,Chemtob87,Scoccola90} have not emphasized the
need for consistent $N_c$-counting. In particular, a {\sl frozen} approximation
in the treatment of the response to electromagnetic perturbations
was made, and the effects of distortions
of the soliton, which occur at a relevant $N_c$-level, were not
included. In this paper we develop consistent $N_c$-counting
rules for $\alpha$ and $\beta$, and evaluate these
observables in a specific model. We point out serious difficulties
arising in the case of the magnetic polarizabilities. Special attention is
paid to the issue of neutron-proton splitting of the polarizabilities, which
has already been reported elsewhere \cite{BBC91nt}.


Throughout this paper we use the so-called
proper (or static) polarizabilities, $\alpha$ and $\beta$,
defined via the energy shift of an object in constant electric, $E$,
or magnetic, $H$, fields \cite{remark:neutral}:
\begin{equation}
\delta{\cal{E}} = - \mbox{$1\over 2$} \alpha E^2 - \mbox{$1\over 2$} \beta H^2.
\label{eq:properpol}
\end{equation}
These are directly related (see Sec. \ref{se:retardation}) to the
coefficients $\overline\alpha$ and $\overline\beta$
encountered in the Compton amplitude
\cite{Ericson73,Baldin,Klein,Petrunkin61,Petrunkin81,Friar:rev,Chemtob87}
\begin{equation}
\overline{\alpha} = \alpha + \Delta\alpha, \;\;
\Delta\alpha = \frac{Q^{2} {\langle r^{2} \rangle}_{E}}{3 M},
\;\; \overline{\beta}  = \beta ,
\label{eq:propercompt}
\end{equation}
where the recoil term $\Delta\alpha$ involves the charge,
$Q$, mass $M$, and electric
mean square radius of the particle.
Although the experimental errors are still substantial \cite{Nathan:talk},
the new measurements
indicate that the proper electric polarizability
is larger for the neutron than for the proton,
\begin{eqnarray}
\alpha_n & = & (12.0 \pm 1.5  \pm 2.0)  \times 10^{-4} fm^{3}, \nonumber \\
\alpha_p & = & (7.2  \pm 1.0  \pm 1.0)  \times 10^{-4} fm^{3},
\label{eq:expalpha}
\end{eqnarray}
and that the magnetic polarizabilities are positive, and equal
within experimental errors:
\begin{eqnarray}
\beta_n & = & (3.1   \mp 1.5  \mp 2.0)  \times 10^{-4} fm^{3}, \nonumber \\
\beta_p & = & (3.4   \mp 1.0  \mp 1.0)  \times 10^{-4} fm^{3}.
\label{eq:expbeta}
\end{eqnarray}

A number of results presented in this paper
are generic to any hedgehog model, e.g. the
$N_c$-counting rules and the role of the
intermediate $\pi - \Delta$ states. Our
numerical results are obtained using the simple
model of Ref. \cite{BirBan8485} (Sec. II of (II)). Since this model
possesses both quark and meson degrees of freedom, it
is in this respect representative of a general class of
hedgehog models. Our calculations
can be repeated in any hedgehog model with minor modifications.

The organization of this article is as follows: In Sec. \ref{se:eandm}
we write down the gauged $\sigma$-model lagrangian, introduce collective
coordinates in the usual way (Sec. {III F} of (II)), and
classify various electromagnetic perturbations according
to the hedgehog symmetries. We discuss the sea-gull and dispersive
contributions to the Compton amplitude (Sec. \ref{se:Compton}), and present
the $N_c$-counting rules (Sec. \ref{se:Nc}).
For the electric polarizabilities we find that the neutron-proton splitting
effect is two powers of $N_c$ suppressed
compared to the average nucleon value:
\begin{eqnarray}
\alpha_{N}   & = & (\alpha_n + \alpha_p)/2 \sim N_c, \nonumber \\
\delta\alpha & = &  \alpha_n - \alpha_p \sim 1/N_c.
\label{eq:ncalpha}
\end{eqnarray}
For the magnetic case the $N_c$-counting
is more complicated (Sec. \ref{se:magnetic}).

In Sec. \ref{se:electric} we obtain numerical results for the electric
polarizabilities. For $\alpha_{N}$, the sea-gull
contribution (Sec. \ref{se:seagull}) is dominant, and the
valence quark effects enter at the level of 10 \%. This is due to the
long-range nature of the pion. Our model prediction
is about a factor of three too large than the experimental number.
We discuss this discrepancy, which is due to two resons.
Firstly, our specific model has a pionic tail which is too strong
(i.e. the value of $g_{\pi NN}$ is too large compared to nature). This enhances
the model prediction by about a factor of 2. A more
fundamental reason is discussed in Sec. \ref{se:chipt}, where we
point out that hedgehog predictions for some observables
largely overestimate results due to the implicit
treatment of the $\Delta$ resonanse as
degenerate with the nucleon.
Our prediction for $\delta\alpha = \alpha_n - \alpha_p$ are less sensitive
on the strength of the pionic tail, and they are not affected
by problems discussed in Sec. \ref{se:chipt}. Numerically, we obtain
$\delta\alpha = 5.4 \times 10^{-4} fm^{3}$ \cite{BBC91nt}, in agreement
with the sign and magnitude given by the experiment.
This splitting arises naturally in hedgehog models when dispersive effects
are taken into account. The effect is dominated by
the distortion of the pionic cloud. We also present a classical
argument why the sign of $\delta\alpha$ is expected to be positive in
models with pionic clouds (Sec. \ref{se:sign}).

In Sec \ref{se:retardation}, we show that the standard retardation correction
(Eq. \ref{eq:propercompt}) strictly holds in our linear response
method.
In Sec. \ref{se:magnetic} we discuss the issue of magnetic polarizability
in hedgehog models. We show that the degeneracy of the $\Delta$ and
nucleon masses in the large-$N_c$ limit precludes the use of linear
response theory to determine $\beta$.  The $N - \Delta$ paramagnetic
contribution, $\beta_{N \Delta}$ (given by
the Born term with the intermediate $\Delta$
state) dominates the $N_c$ behavior of $\beta$, and leads to
\begin{equation}
\beta_{N} = (\beta_n + \beta_p)/2 \sim N_c^3,
\label{eq:ncbeta}
\end{equation}
In our view, this invalidates the claims of Refs.
\cite{Nyman84,Chemtob87,Scoccola90} that
the diamagnetic sea-gull interaction cancels the paramagnetic $\Delta$ term,
since these terms occur at different $N_c$-levels
In Sec. \ref{se:magndisp}
we also show the relevance of
dispersive contributions in the calculation of the magnetic
polarizability.

In Sec. \ref{se:chipt} we compare our results
to predictions of the chiral perturbation
theory ($\chi PT$). We find, that in the chiral limit the hedgehog predictions
and
the $\chi PT$ predictions for $\alpha$ and $\beta$
agree up to a factor which can be attributed to the
role of the $\Delta$ in pionic loops. We discuss interesting physical
implications of this issue, and point out that while the naive hedgehog
predictions
overestimate the role of the $\Delta$ resonance in hadronic loops,
the naive approach to $\chi PT$
drops these important contributions altogether.
We discuss the effects of finite
$N - \Delta$ mass splitting in these approaches, and propose how
the hedgehog and $\chi PT$ results should be corrected.

In Sec. \ref{se:substructure} we study the effects
of the pionic substructure on nucleon polarizabilities. These effects are
introduced
via non-minimal substitution terms ${\cal L}_9$ and ${\cal L}_{10}$
\cite{Holstein,Gasser} in the effective
lagrangian. At the mean-field level, we obtain simple expressions
involving pion polarizability and pion mean squared radius. Numerically, these
pion structure contributions enter at the level
of $\sim 1 - 2 \times 10^{-4} fm^{3}$, which is a small but non-negligible
effect.

In Sec. \ref{se:othermodels} we make a few comments relevant to
calculations in other hedgehog models.

Throughout the paper we use the units $\hbar = 1$, $c = 1$, $e^2/(\hbar c) =
1/137$.

\section{Electromagnetic perturbations}
\label{se:eandm}

Our basic methods of treatment of external perturbations in hedgehog models
are described in (II). The approach developed in (II)
is a rather straightforward adaptation
of the RPA method of traditional many-body physics. The diference with
the traditional nuclear physics case is the presence of mesonic degrees
of freedom, relativistic dynamics, and special symmetries (Sec. II B in (II)).

Polarizabilities are defined as second-order energy shifts
due to (static) external electromagnetic
fields, Eq. (\ref{eq:properpol}), or, equivalently,
via Compton amplitude (Sec. \ref{se:Compton}). They arise
from two types of terms: sea-gulls, which
are due to quadratic terms in $E$ or $B$ in the lagrangian,
and from dispersive terms.
Sea-gull effects result in local expressions involving
the pion mean field \cite{Nyman84,Chemtob87,Scoccola90},
while the dispersive terms
are given by the usual second-order perturbation theory expressions, which
result from linear perturbations in the lagrangian. These dispersive effects
distort the soliton. They are particularly important in
the neutron-proton splitting effects, as well as in the magnetic response,
and have
not been considered in earlier works. More importantly,
these effects enter at the
same $N_c$-level as the sea-gull terms, and thus should be included in order to
comply with the basic organizational principle
of the hedgehog approach, namely, the
$1/N_c$-perturbation expansion. The dispersive terms in the
Compton amplitude correspond, in our approach,  to diagrams in which
the propagator between the two photon insertions is the RPA (or linear
response)
propagator (Fig. \ref{fi:rpa}). Relevant expressions are obtained using
equations-of-motion
method, which amounts to solving linear differential equations with
potentials and driving terms determined by the solitonic profile. In addition,
electromagnetic perturbations are carried in the presence of cranking, which
ensures projection on states with good quantum numbers (Sec. II G in (II)).

\subsection{Lagrangian in presence of electromagnetic interactions}
\label{se:lagr}

The first step is to identify perturbations resulting from coupling
to electromagnetic interactions, which
are introduced by gauging the $\sigma$-model
lagrangian (II.2.1):
\begin{eqnarray}
 {\cal L} &=& \bar{\psi} \left [ \dot{\imath} \overlay{\slash}{\partial}
   + g \left ( \sigma+\dot{\imath} {\gamma_{5}} \mbox{\boldmath $\tau$} \cdot
\mbox{\boldmath $\pi$} \right ) \right ] \psi \nonumber \\
    &+&  \mbox{$1\over 2$} (\partial{}^{\mu} \sigma)^{2}
  + \mbox{$1\over 2$} (\partial{}^{\mu} \mbox{\boldmath $\pi$})^{2}
   - U \left ( \sigma, \mbox{\boldmath $\pi$} \right ).
\label{eq:GML}
\end{eqnarray}
Through this procedure one generates a
dispersive interaction, as well as a covariantizing sea-gull term:
\FL
\begin{eqnarray}
&&{\cal L}_{disp.}   =  - e A^{\mu}
  \left ( \overline{\psi} \left ( \frac{1}{2 N_c} + \mbox{$1\over 2$}
\tau_{3} \right ) \gamma_{\mu} \psi
 + {\left ( \mbox{\boldmath $\pi$} \times \partial{}_{\mu}
   \mbox{\boldmath $\pi$} \right )}_{3} \right ), \nonumber \\
&& {\cal L}_{sea-gull}   =  \mbox{$1\over 2$} e^{2} A^{\mu} A_{\mu}
  \left ( \mbox{\boldmath $\pi$}_{1}^{2} +
   \mbox{\boldmath $\pi$}_{2}^{2} \right ) ,
\label{eq:EM}
\end{eqnarray}
where $\psi$ is the quark field, $\mbox{\boldmath $\pi$}$ is
the pion field, and $A^{\mu}$ is the photon field. It is
hoped that this minimal substitution procedure
gives the bulk of electromagnetic interactions in our
effective hadronic lagrangian, however, non-minimal substitution terms
may a priori play an important role. In
Section \ref{se:substructure} we discuss the role of simple non-minimal
substitution terms in effective lagrangians, and find that their
effects enter (in the electric polarizability) at the level of 10 \%, compared
to the terms resulting from (\ref{eq:EM}).

To identify the appropriate perturbations, we make a transformation
of the full lagrangian (II.2.1, \ref{eq:EM}) to the isorotating
frame, according to Eqs. (II.3.22). This is done because
we are interested in the linear response of the nucleon, which
is ``projected out'' of the hedgehog (see Sec. III in (II) for
details of projection via cranking, and linear response in presence of
cranking). For a more compact notation, we also replace the quark field
bilinears by $N_c$ times the  bilinears of the valence quark, $q$.
The Lorentz gauge is used, and
$A^{0} = - \mbox{\boldmath $r$} \cdot \mbox{\boldmath $E$}$,
$\mbox{\boldmath $A$} = - \mbox{$1\over 2$} \mbox{\boldmath $r$}
\times \mbox{\boldmath $B$}$. We obtain
\FL
\begin{eqnarray}
{\cal L}  &\rightarrow&  {\cal L}^{0} + {\cal L}_{\lambda}^{1}
 + {\cal L}_{\lambda}^{2} + {\cal L}_{{E}^{0}}^{1}
+ {\cal L}_{{E}^{1}}^{1}
 + {\cal L}_{E^{1}}^{2} + {\cal L}_{\lambda E^{1}}^{2}
\nonumber \\
 &+& {\cal L}_{{B}^{0}}^{1} + {\cal L}_{{B}^{1}}^{1}
 + {\cal L}_{B^{1}}^{2} ,
 \label{eq:ltrans} \\
\nonumber \\
{\cal L}^{0}  &=& N_c \,\overline{q}
\left ( \dot{\imath} \overlay{\slash}{\partial}
   + g \left ( \sigma+\dot{\imath} {\gamma_{5}}
\mbox{\boldmath $\tau$} \cdot \mbox{\boldmath $\pi$} \right )
\right ) q \nonumber \\
&+& \mbox{$1\over 2$} ({\partial^\mu} \sigma)^{2}
+ \mbox{$1\over 2$} ({\partial^\mu} \mbox{\boldmath $\pi$})^{2}
   - U \left ( \sigma, \mbox{\boldmath $\pi$} \right ) ,
\label{eq:l0} \\
\nonumber \\
{\cal L}_{\lambda}^{1}  &=&  - N_c \,{q}^{\dagger}
 \left ( \mbox{$1\over 2$} \mbox{\boldmath $\lambda$}
\cdot \mbox{\boldmath $\tau$} \right ) q
 - \mbox{\boldmath $\lambda$} \cdot
\left ( \mbox{\boldmath $\pi$} \times
\dot{\mbox{\boldmath $\pi$}} \right ) ,
\label{eq:llam1} \\
{\cal L}_{\lambda}^{2}  &=&  \mbox{$1\over 2$}
{\left ( \mbox{\boldmath $\lambda$} \times
\mbox{\boldmath $\pi$} \right )}^{2} ,
\label{eq:llam2} \\
\nonumber \\
{\cal L}_{{E}^{0}}^{1}  &=&  e \,\mbox{\boldmath $r$}
\cdot \mbox{\boldmath $E$} \,
 \mbox{$1\over 2$} {q}^{\dagger} q ,
\label{eq:lE0} \\
{\cal L}_{{E}^{1}}^{1}  &=&  e \,\mbox{\boldmath $r$}
\cdot \mbox{\boldmath $E$}
 \left ( N_c \,{q}^{\dagger}
 \mbox{$1\over 2$} \mbox{\boldmath $c$} \cdot
\mbox{\boldmath $\tau$}  q
 + \mbox{\boldmath $c$} \cdot \left
( \mbox{\boldmath $\pi$} \times
 \dot{\mbox{\boldmath $\pi$}} \right ) \right ),
\label{eq:lE11} \\
{\cal L}_{E^{1}}^{2}  &=&  \mbox{$1\over 2$} e^2
(\mbox{\boldmath $r$} \cdot \mbox{\boldmath $E$})^{2}
{\left ( \mbox{\boldmath $c$} \times \mbox{\boldmath $\pi$} \right )}^{2} ,
\label{eq:lE12} \\
{\cal L}_{\lambda E^{1}}^{2}  &=&  - e \,\mbox{\boldmath $r$}
\cdot \mbox{\boldmath $E$} \,
 {\left ( \mbox{\boldmath $c$} \times {\mbox{\boldmath $\pi$}} \right )}
\cdot {\left ( \mbox{\boldmath $\lambda$}
\times {\mbox{\boldmath $\pi$}} \right )} ,
\label{eq:llamE1} \\
\nonumber \\
{\cal L}_{{B}^{0}}^{1}  &=&   - \mbox{$1\over 4$} e \,(\mbox{\boldmath
$r$} \times \mbox{\boldmath $B$})^i
 {q}^{\dagger} \alpha_i q ,
\label{eq:lB0} \\
{\cal L}_{{B}^{1}}^{1}  &=&  - \mbox{$1\over 2$} e \,(\mbox{\boldmath
$r$} \times \mbox{\boldmath $B$})^i
 \bigl ( N_c \,{q}^{\dagger}
  \mbox{$1\over 2$} \mbox{\boldmath $c$} \cdot \mbox{\boldmath $\tau$}
\alpha_i q \nonumber \\
 &-&  \mbox{\boldmath $c$} \cdot ( \mbox{\boldmath $\pi$} \times
   \nabla_i \mbox{\boldmath $\pi$} ) \bigr ) ,
\label{eq:lB11} \\
{\cal L}_{B^{1}}^{2}  &=&  - \frac{1}{8} e^2 \,(\mbox{\boldmath $r$}
\times \mbox{\boldmath $B$})^2
{\left ( \mbox{\boldmath $c$} \times \mbox{\boldmath $\pi$} \right )}^{2} ,
\label{eq:lB12}
\end{eqnarray}
where $\mbox{\boldmath $\lambda$}$ is the
cranking velocity, and the collective vector
$\mbox{\boldmath $c$} = \mbox{$1\over 2$} Tr[\tau_3 B
\mbox{\boldmath $\tau$} B^{\dagger}]$ is defined in App. B in (II).
In the above equations, subscript $\lambda$ refers to term arising upon
cranking, $E^0$ and $E^1$ denote isoscalar and isovector electric
perturbations, and $B^0$ and $B^1$ denote isoscalar and isovector magnetic
perturbations. Indices $i$ in Eqs. (\ref{eq:lB0} - \ref{eq:lB11})
are spatial Cartesian indices.
Superscript ${}^1$ denotes dispersive terms, which lead to
linear shifts in fields according to perturbation theory. Superscript ${}^2$
denotes terms which are quadratic in perturbations. They include
a term quadratic in $\lambda$ (Eq. \ref{eq:llam2}), sea-gull terms
(Eqs. \ref{eq:lE12},\ref{eq:lB12}), and a mixed term (Eq. \ref{eq:llamE1}),
which will play an essential role in the splitting of the
neutron and proton polarizabilities. The classification of various
dispersive terms, as well as the explicit forms of the
corresponding sources arising in
linear response equations, are given in Table \ref{tab:kpr}. The
grand spin (sum of spin and isospin) is denoted by $K$, parity is
denoted by ${\cal P}$, and the grand-reversal symmetry by ${\cal R}$
(see Sec. {II B} in (II) for details).

\subsection{Compton amplitude and polarizabilities}
\label{se:Compton}

The Compton amplitude corresponding to
lagrangian (\ref{eq:ltrans}) reflects the presence of both dispersive
and sea-gull terms. It can be written as \cite{Chemtob87}
\FL
\begin{eqnarray}
&&\tilde{M}_{\mu \nu}(p,q) = M_{\mu \nu}(p,q) + S_{\mu \nu}(q) ,
\nonumber \\
&&M_{\mu \nu}(p,q) = \nonumber \\
& &\dot{\imath} \int d^4 x \, e^{\dot{\imath} q \cdot x}
  \langle N(p') |T \left ( J^{e.m.}_{\mu}(x)
  J^{e.m.}_{\nu}(0) \right ) | N(p) \rangle , \nonumber \\
\label{eq:Compton}
\end{eqnarray}
where $M_{\mu \nu}$ is the dispersive $T$-product part, and $S_{\mu \nu}$ is
the sea-gull part. Correspondingly, the polarizabilities have dispersive
as well as sea-gull parts, and can be written as \cite{Chemtob87}
\FL
\begin{eqnarray}
\alpha &=& \alpha_{disp.} + \alpha_{sea-gull} , \;\;\;
    \beta  = \beta_{disp.}  + \beta_{sea-gull} , \nonumber \\
\alpha_{disp.} &=& 2 \sum_{b'}
\frac{| \langle N| \int d^3 x \,
\widehat{\mbox{\boldmath $E$}} \cdot \mbox{\boldmath $r$}
J^{e.m.}_{0}(\mbox{\boldmath $r$})
|b' \rangle |^2}{E_{b'} - E_N} , \nonumber \\
\alpha_{sea-gull} &=&  \langle N|\int d^3 x \,
(\widehat{\mbox{\boldmath $E$}} \cdot
\mbox{\boldmath $r$})^2 S_{00}|N \rangle , \nonumber \\
\beta_{disp.} &=& 2 \sum_{b'}
\frac{| \langle N| \int d^3 x \,
\mbox{$1\over 2$} \widehat{\mbox{\boldmath $B$}} \cdot
(\mbox{\boldmath $r$} \times \mbox{\boldmath $J$}^{e.m.}
(\mbox{\boldmath $r$})) |b' \rangle |^2}{E_{b'} - E_N} , \nonumber \\
\beta_{sea-gull} &=&  \mbox{$1\over 4$} \langle N|\int d^3 x \,
\epsilon_{mni} \widehat{B}^m r^n
\epsilon_{klj} \widehat{B}^k r^l S_{ij}|N \rangle .  \nonumber \\
\label{eq:polardecomp}
\end{eqnarray}
Hats denote unit vectors in the direction
of $E$ or $B$ fields, and $|b' \rangle$ is
an intermediate state with energy $E_{b'}$.

The sea-gull parts of expressions (\ref{eq:polardecomp}) can be readily
identified from Eq. (\ref{eq:lE12},\ref{eq:lB12}). One gets
\begin{eqnarray}
S_{00} &=& e^2 {\left ( \mbox{\boldmath $c$} \times
\mbox{\boldmath $\pi$} \right )}^{2}, \nonumber \\
S_{ij} &=& - e^2\delta_{ij} {\left ( \mbox{\boldmath $c$}
\times \mbox{\boldmath $\pi$} \right )}^{2}.
\label{eq:sgch}
\end{eqnarray}
Evaluating these in the collective nucleon state one
obtains (App. B in (II), \cite{Chemtob87,Scoccola90})
\begin{eqnarray}
\alpha_{sea-gull} = - 2 \beta_{sea-gull} =
\frac{8 \pi e^2}{9} \int dr \, r^4 \pi_h^2 ,
\label{eq:polarsg}
\end{eqnarray}
where $\pi_h$ is the hedgehog pion field profile.
In Eq. (\ref{eq:polarsg}) we recognize
the hedgehog model relation between the seagull
contributions to the electric and magnetic
polarizabilities \cite{Chemtob87,Scoccola90}.
The radial integral in Eq. (\ref{eq:polarsg}) is quadratic
in the pion field. In the Skyrme model additional
terms arise due to higher-order terms in the lagrangian
\cite{Nyman84,Chemtob87,Scoccola90}.

\subsection{Linear response}
\label{se:linresp}

In our treatment of dispersive contributions,
the intermediate states $|b' \rangle$
in Eq. (\ref{eq:polardecomp}) correspond to RPA excited states,
and the energy denominators involve energies of these
excitations (Fig. \ref{fi:rpa}).
These contributions can be written in the general form given in
Eqs. (II.3.27). The electric and magnetic dispersive
polarizabilities result from either isoscalar
dipole or isovector dipole transitions,
and will be labeled by $E^0$ and $E^1$, or $B^0$ and $B^1$, respectively.
These dispersive contributions, as well as
the sea-gull contributions (\ref{eq:polarsg})
contribute equal amounts to the neutron and the proton.
This is because the resulting collective
operators in Eqs. (II.3.27) are isoscalar (bilinear in the
collective vector $\mbox{\boldmath $c$}$),
and matrix elements are equal
for the proton and the neutron. The neutron-proton splitting
effects arise from the mechanism
described in Sec. {III G} in (II). This can be briefly
described in the following way:
We perform electromagnetic perturbations not on hedgehogs, but on nucleons.
These are obtained from hedgehogs via projection (cranking), which in turn
may also be viewed as linear response. It is sufficient to work to linear
order in the cranking velocity,
$\mbox{\boldmath $\lambda$}$, and we can write down
symbolically
\begin{equation}
|N \rangle = (1 + {\cal G}{\cal V}_{cr}) |H \rangle,
\label{eq:crnh}
\end{equation}
where ${\cal G}$ is the RPA propagator, ${\cal V}_{cr}$ is the cranking
interaction, and $|H \rangle$ is the hedgehog state.
Using these ``cranked'' states in our perturbation
theory leads to expressions of the form of
Eqs. (II.3.29), with one cranking, and two
electromagnetic interactions. As discussed in Sec. {III G} in (II),
the $K^{\cal P}$ numbers
of interactions in Eqs. (II.3.29) have to be
additive to $K^{\cal P} = 0^{+}$, since the
hedgehog has $K^{\cal P} = 0^{+}$. From
Table \ref{tab:kpr} we can see that
we can compose cranking, one isoscalar electric (magnetic),
and one isovector electric (magnetic) interaction
to $K^{\cal P} = 0^{+}$. Since ${\cal V}_{cr}$
carries a collective operator $\lambda$,
the appropriate collective operator is
isovector, and neutron-proton splitting
of polarizabilities is generated.

{}From a slightly different but equivalent point of view,
we may understand the isospin effect in electric polarizabilities
by inspecting the term (\ref{eq:llamE1}).
It is linear in cranking, and linear in isovector
electric ($E^1$) perturbation. It leads to
the following contribution to $\alpha$:
\FL
\begin{equation}
\alpha_{\lambda}^{mes.} =
- e \, \langle N|\int d^{3} r \,
(\widehat{\mbox{\boldmath $E$}} \cdot \mbox{\boldmath $r$})
  [( \mbox{\boldmath $c$} \times \mbox{\boldmath $\pi$}) \cdot
( \mbox{\boldmath $\lambda$} \times
\mbox{\boldmath $\pi$} )] |N \rangle .
\label{eq:splitmes1}
\end{equation}
If the pion field in the above equation
were taken to be just the hedgehog profile,
then $\alpha_{\lambda}^{mes.}$ would vanish by parity. However,
the isoscalar electric perturbation (Section \ref{se:E0}) distorts the
meson fields, and $\mbox{\boldmath $\pi$} =
\mbox{\boldmath $\pi$}_h + \delta \mbox{\boldmath $\pi$}_{E^0}$, where
$\delta \mbox{\boldmath $\pi$}_{E^0}$ has  $S$- and $D$-wave
components. As a result, the integral is non-zero.
The corresponding collective operator is
$\mbox{\boldmath $c$} \cdot \mbox{\boldmath $\lambda$} = I_3 / \Theta$,
where $\Theta$ is the moment of inertia (see App. B in (II)), and we obtain
\FL
\begin{equation}
\alpha_{\lambda}^{mes.} =
- e \,\frac{I_3}{\Theta} \, 8/3 \, \int dr \,
(\widehat{\mbox{\boldmath $E$}} \cdot \mbox{\boldmath $r$})
  (\mbox{\boldmath $\pi$}_h \cdot \delta \mbox{\boldmath $\pi$}_{E^0} ), .
\label{eq:splitmes2}
\end{equation}
{}From the quark parts of Eqs. (\ref{eq:ltrans}) we get
\FL
\begin{eqnarray}
\alpha_{\lambda}^{quark} &=&
2 \langle N| \int d^{3} r \, [ - \delta q_{E^{0}}^\dagger
\mbox{$1\over 2$} \mbox{\boldmath $\lambda$}
\cdot \mbox{\boldmath $\tau$} \delta q_{E^{1}}
+ \delta q_{E^{1}}^\dagger \frac{e \widehat{\mbox{\boldmath $E$}}
\cdot \mbox{\boldmath $r$}}{2 N_c}
\delta q_{\lambda} \nonumber \\
  &+& \delta q_{E^{0}}^\dagger \mbox{$1\over 2$} e
\widehat{\mbox{\boldmath $E$}}
\cdot \mbox{\boldmath $r$} \mbox{\boldmath $c$}
\cdot \mbox{\boldmath $\tau$}
\delta q_{\lambda} ] |N \rangle + h.c. ,
  \label{eq:splitq}
\end{eqnarray}
where various terms in the interaction are sandwiched with shifts corresponding
to
other interactions. Eqs. (\ref{eq:splitmes2},\ref{eq:splitq}) can be
straightforwardly obtained from Eq. (II.3.29).
Terms (\ref{eq:splitmes2},\ref{eq:splitq}) are proportional
to the nucleon isospin, $I_3$, thus are
responsible for splitting of the neutron and
proton electric polarizabilities, $\delta \alpha$.

For the splitting of magnetic polarizabilities, mesons do not
contribute at the linear-response level, since there
is no mesonic analog of the term  (\ref{eq:llamE1}) in
lagrangian (\ref{eq:ltrans}). Only the magnetic analog of the quark part
(Eq. \ref{eq:splitq}) is present. However, the linear response
calculation of the magnetic polarizability encounters fundamental
problems, which will be discussed in Sec. \ref{se:magnetic}.

\subsection{$N_c$ - counting}
\label{se:Nc}

In Eqs. \ref{eq:ltrans} we list explicitly
occurrences of $N_c$. We also recall \cite{witten:nc}
that in the large-$N_c$ limit the meson fields
scale as $\sqrt{N_c}$, the
quark fields scale as $1$, and the moment of
inertia scales as $N_c$.  Note that the
sources for isoscalar electromagnetic
perturbations are one power down compared to the corresponding
sources for the isovector interactions. We easily
arrive at the following $N_c$-rules for electric polarizabilities:
\FL
\begin{equation}
\alpha_{sea-gull} \sim N_c , \;\; \alpha_{E^0} \sim 1/N_c , \;\;
\alpha_{E^1} \sim N_c , \;\; \alpha_{\lambda} \sim 1/N_c .
\label{eq:Nc}
\end{equation}
This leads directly to the following rules for the nucleon polarizabilities:
\begin{eqnarray}
\alpha_N &=& \frac{\alpha_n + \alpha_p}{2} = \alpha_{sea-gull} + \alpha_{E^1}
\sim N_c, \nonumber \\
\delta \alpha &=& \alpha_n - \alpha_p = \alpha_{\lambda} \sim 1/N_c.
\label{eq:genericnc}
\end{eqnarray}
Note, that since the $\alpha_{E^0}$ part of the dispersive term contributes
to $\alpha_N$ at a subleading level, it should be dropped according to
$N_c$-rules. Other physical effects affect our results at this level, e.g.
centrifugal stretching, center-of-mass corrections to the soliton mass,
$N_c$-suppressed terms in the effective lagrangian, etc.
Numerically, the $\alpha_{E^0}$ contribution is negligible,
confiring the validity of $N_c$-counting for polarizabilities.

Naively, one would write down expressions analogous
to Eqs. (\ref{eq:genericnc}) for the magnetic
polarizability, $\beta$. We will show in Sec. \ref{se:magnetic}
that this is not correct.

\section{Electric polarizability}
\label{se:electric}

In this section we present our numerical
results for $\alpha$. Numerical methods,
explicit forms of the differential equations,
and other details are given in (II).
The model parameters for our numerical calculations are taken from
Refs. \cite{BirBan8485,CB86}: $F_{\pi} = 93 MeV$, $g F_{\pi} = 500 MeV$,
$m_\pi = 139.6 MeV$, and $m_\sigma = 1200 MeV$.

\subsection{Sea-gull contribution}
\label{se:seagull}

Because the pion field is long ranged, the nucleon
electric polarizability, $\alpha_N$, is dominated by the pionic
sea-gull contribution (\ref{eq:polarsg}). Numerically,
in our model we get
\begin{equation}
\alpha_{sea-gull} = 28 \times 10^{-4} fm^3 , \label{eq:sgnum}
\end{equation}
which is roughly three times larger than the experimental
value in (\ref{eq:expalpha}). There are
two basic reasons for this discrepancy.
The first one is discussed below, and the other in Sec. \ref{se:chipt}.
Figure (\ref{fi:radsg}) shows
the radial density of the integrand of (\ref{eq:polarsg}) (solid line).
It is clear that the sea-gull term
acquires most of its value from the asymptotic region, $r > 1 fm$.
Recall, that the asymptotic behavior of the pion
tail in hedgehog models is determined by the pion-nucleon coupling
constant, $g_{\pi NN}$ \cite{Adkins84,CB86}.
Using the Goldberger-Treiman relation, we
can write down
\begin{equation}
\mbox{\boldmath $\pi$}_h^{asympt.} = (3 g_A)
 /(8 \pi F_{\pi}) \hat{\mbox{\boldmath $x$}}
         (m_{\pi} + 1/r) exp(-m_{\pi} r) /r.
\label{eq:pionasympt}
\end{equation}
Our model predicts $g_A = 1.86$ \cite{CB86},
hence the tail contribution is overestimated
by a factor of $(g_A^{model}/g_A^{exp.})^2 \sim 2$. The dotted line
in Fig. (\ref{fi:radsg}) shows the integrand of Eq. (\ref{eq:polarsg})
with the pion field having the form
(\ref{eq:pionasympt}) in the whole region of $r$.
This corresponds to the chiral limit case,
and will be discussed in detail in
Sec. \ref{se:chipt}.
We can see from Fig. (\ref{fi:radsg}) that the size of
$\alpha_{sea-gull}$ is controlled by
how fast the solid line departs from the dashed line at lower values of $r$.
In other words, the results
are sensitive to the profile of the pion field in the intermediate region
of about $1 fm$. Note, that $\pi_h$ enters the expression (\ref{eq:polarsg})
quadratically. Therefore, we expect substantial model sensitivity in
hedgehog predictions of $\alpha_{sea-gull}$, e.g. models which have
a suppressed pion field, such as hedgehog
models with confinement \cite{BBC:rev,MKB:conf,BrLj,Duck},
or models with vector mesons, e.g. \cite{BKY,BB8586},
are expected to predict lower values for
$\alpha_{sea-gull}$. We stress that to compare fairly the model
predictions with experiment, the model
should predict correctly the quantity
$g_A / F_{\pi}$, which enters the asymptotic
form  (\ref{eq:pionasympt}).

In Sec. \ref{se:chipt} we show how the sea-gull contribution is
additionally suppressed when $N - \Delta$
mass splitting effect is taken into account.
This results in another factor of $\sim 2$ reduction. With these
corrections we note that it is possible to put the sea-gull
contribution in the right
experimental range. The model uncertainty, however, is big.

\subsection{Isovector electric perturbation}
\label{se:E1}

In Sec. \ref{se:Nc} we have shown that
the dispersive electric isovector ($E^1$)
contribution to polarizability,
$\alpha_{E^1}$, is of the order $N_c$, the same
order as for $\alpha_{sea-gull}$.
The $E^{1}$ perturbation in lagrangian
(\ref{eq:lE11})
has $K^{\cal PR} = 0^{--}, 1^{--}$ and
$2^{--}$ (Table \ref{tab:kpr}).  Since it is
odd under grand-reversal ${\cal R}$, only the
valence quark fields acquire shifts
(Sec. III A in (II)), and we solve equations of the
form of Eqs. (II.3.7):
\begin{equation}
(h - \varepsilon) q_{E^1} =
 \mbox{$1\over 2$} N_c e \,\mbox{\boldmath $r$}
\cdot \widehat{\mbox{\boldmath $E$}} \,
\mbox{\boldmath $c$} \cdot \mbox{\boldmath $\tau$} q_h .
\label{eq:E1eq}
\end{equation}
These equations are decomposed into $K = 0, 1$
and $2$ components, as described in
App. A in (II), and solved numerically. The
corresponding polarizability is calculated
according to the expression
\begin{equation}
\alpha_{E^1} = - 2 N_c e \, \langle N| \int d^3x
q^{\dagger}_{E_1} \mbox{\boldmath $r$} \cdot
\widehat{\mbox{\boldmath $E$}} \mbox{\boldmath $c$} \cdot
\mbox{\boldmath $\tau$} q_h |N \rangle,
\label{eq:E1pol}
\end{equation}
where $|N \rangle$ is the nucleon collective state (App. B in (II)).
Note that with our definition, the spinor $q_{E^1}$
carries the collective variable
$\mbox{\boldmath $c$}$, such that
Eq. (\ref{eq:E1pol}) contains a matrix
element of a collective operator quadratic in
$\mbox{\boldmath $c$}$, similarly to the case of
the sea-gull part in Sec. \ref{se:Compton}.

Numerically, we get in our model
\begin{equation}
\alpha_{E^1} = 3.49 \times 10^{-4} fm^3,
\label{eq:alphaE1num}
\end{equation}
with $K = 0, 1$, and $2$ components of $q_{E_1}$
contributing $-12 \%$, $26 \%$, and $86 \%$,
respectively. The dashed line in Fig. \ref{fi:radsg}
depicts the radial density of
$\alpha_{E^1}$. Because the quark mass
in our model is $500 MeV$, the purely
quark $\alpha_{E^1}$ contribution
is strongly suppressed compared to the
pion sea-gull contribution. However, the value
of $\alpha_{E^1}$ is non-negligible compared
to experimental numbers (Table \ref{tab:alpha}),
and quark effects are substantial.

In models with vector mesons \cite{BKY,BB8586},
the $E^1$ perturbation involves
shifts in the $\rho$ and $\omega$ mesons, and
these effects must be included
in calculations of the electric polarizability
\cite{remark:crankvec}. They enter at the leading-$N_c$
level, and should contribute comparably to the valence quark effects.

\subsection{Isoscalar electric perturbation}
\label{se:E0}

Now we turn to a more difficult case of
the isoscalar electric perturbation, $E^0$,
which has $K^{\cal PR} = 1^{-+}$. These are quantum numbers of the
translational zero mode, and, as explained in Sec. III A in (II),
we have to deal with full RPA
equations of the form (II.3.3). The appearance
of the zero mode is easy to understand. The hedgehog
soliton possesses isoscalar electric charge,
$Q^{I=0}=\mbox{$1\over 2$} e$. Thus, in a constant
electric field, the soliton accelerates
in the direction of $\mbox{\boldmath $E$}$
--- the translational mode is excited.
In addition to this zero-mode motion, the soliton
is getting distorted (as viewed from the center of mass).
Below we show that the excitation of the translational motion
corresponds to the Thompson limit of the
Compton scattering, and the deformation corresponds to
the polarizability. In order to separate the zero
mode from the physical modes, instead
of a constant electric field we consider
a slowly time-dependent field of the form
\begin{equation}
E(t) = \mbox{$1\over 2$} \left ( E_{0}
e^{- \dot{\imath} \omega t} +
E^{*}_{0} e^{\dot{\imath} \omega t} \right )
\label{eq:Etimedep}
\end{equation}
The resulting sources in Eq. (II.3.3) are
\begin{eqnarray}
& & j_X = j_Y = \mbox{$1\over 2$}\, e \,
N_c^{-1} (\mbox{\boldmath $r$} \cdot
\widehat{\mbox{\boldmath $E$}}) q_h, \nonumber \\
& & j_Z = j_P = 0 .
\label{eq:alsources}
\end{eqnarray}
Using the technique described in Sec. III B in (II),
we find that the total isoscalar electric
polarizability consist of the
translational zero-mode part, and
a ``physical'' part:
\begin{equation}
\alpha_{E^0}^{tot.} = \alpha_{E^0}^{zero} + \alpha_{E^0}^{phys.},
\label{eq:alphatot}
\end{equation}
where the zero-mode part diverges in the limit $\omega \rightarrow 0$,
\begin{equation}
\alpha_{E^0}^{zero} = - \frac{(Q^{I=0})^2}{M \omega^2},
\label{eq:alphazero}
\end{equation}
and the physical part, $\alpha_{E^0}^{phys.}$,
has a finite $\omega \rightarrow 0$ limit.
The quantity $M$ in Eq. (\ref{eq:alphazero}) is the soliton mass.
In the expression for the forward
Compton scattering amplitude, $\alpha$ is multiplied
by $\omega^{2}$ \cite{Petrunkin61,Ericson73,Friar:rev}.
We note immediately that
$\omega^{2} \alpha_{E^0}^{zero} = - \frac{(Q^{I=0})^2}{M}$
is just the Thompson term in scattering of a particle with
charge $Q^{I=0}$ and mass $M$. Thus, the zero-mode part is responsible
for the Thompson limit. Since RPA leads
to small fluctuation equations of motion,
analogously to the case of classical
physics \cite{Ericson73}, it is clear
it leads to the correct Thompson limit.

In the case of the $E^1$ perturbation, the
quantum numbers precluded excitation of a zero-mode
on top of the hedgehog solution. This is
a manifestation of the fact that the hedgehog
does not have any isovector electric
charge, $Q^{I=1} = 0$. This charge arises only
upon cranking, but the organization
of our perturbation theory is such, that
we treat cranking, $E^0$ and $E^1$
perturbations separately (Sec. III G and V in (II)).
If linear response were performed
on a cranked soliton, then the full
charge $Q = Q^{I=0} + Q^{I=1}$ would appear in Thompson scattering.

Earlier works \cite{Chemtob87,Scoccola90}
on electric polarizabilities
in hedgehog models did not take into
account the effects of zero modes. In fact, this is justified by the
$1/N_c$ expansion.  The $E^0$ contribution is suppressed by two powers
of $N_c$ compared to the $E^1$
contribution (Eq. (\ref{eq:Nc}). Therefore,
if one is interested in the leading
$N_c$ behavior of $\alpha_N$, it is sufficient
to consider the isovector electric perturbation only, where
the issue of the translational zero mode does not arise.

Since we are interested in the splitting
of the neutron and proton polarizabilities,
we have to calculate the
quantities $\alpha_{\lambda}^{mes.}$
(Eq. (\ref{eq:splitmes2})) and $\alpha_{\lambda}^{quark}$
(Eq. (\ref{eq:splitq})). Thus, we have to find the shifts
in the fields due to the isoscalar perturbation,
$E^0$. We need to extract the ``physical''
parts of the solution. The numerical procedure
has been described in Sec. III C in (II).
Here we only remark, that very good numerical
accuracy is necessary in order to
separate the zero mode from the physical mode.
This is because at small values of
$\omega$ the solution can be written as
\begin{equation}
\xi = \frac{\dot{\dot{\imath}ath} Q_0}{{\cal M} \omega^2} \xi_0
     - \frac{Q_0}{{\cal M} \omega}
\xi_1 + \xi_{phys} + {\cal O}(\omega),
\label{eq:xi}
\end{equation}
where $\xi_0$ is the translational
mode, $\xi_1$ is the conjugated mode
(``boost mode''), and $\xi_{phys.}$
is the physical mode which we want
to extract (Sec. III B in (II)).
Solutions of equations (II.3.3) in the limit
$\omega \rightarrow 0$ give full $\xi$,
from which we subtract the pieces
divergent in $\omega$. This can
easily be done, since we know the exact forms of
$\xi_0$ and $\xi_1$ (App. C in (II)).
The presence of the zero mode provides a useful
algebraic check of equations from App. A in (II), since the
divergent parts of Eq. (\ref{eq:xi}) have known coefficients.

\subsection{Neutron-proton splitting of electric polarizabilities}

Having solved the linear response equations for the $E^0$ perturbation,
we use $q_{E^0}^{phys.} = X^{phys.}+Y^{phys}$
and $\phi^{phys.} = 2 Z^{phys.}$
(App. A in (II))
in equations (\ref{eq:splitmes2}, \ref{eq:splitq}). The dominant part
to $\alpha_{\lambda}$ comes from the pionic contribution. The pion shift
has $S$- and $D$-wave components (Table \ref{tab:kpr}), which we denote by
$\pi_S$ and $\pi_D$, respectively.  The explicit
expression for the mesonic contribution of
$\delta\alpha^{mes}$ has the form
\begin{equation}
\delta\alpha^{mes} =
- \frac{16 \, e}{3 \Theta}  \sqrt{\frac{4 \pi}{3}}
\int dr\,r^3 \pi_h ( \sqrt{1/3} \pi_S  - \sqrt{2/3} \pi_D .
\label{eq:allampi}
\end{equation}
The quark contribution is obtained analogously from Eq. (\ref{eq:splitq}).
The numerical results give
\begin{equation}
\delta\alpha = - 5.6 \times 10^{-4} fm^3 ,
\label{eq:allamnum}
\end{equation}
with the quarks carrying $- 5 \%$ of the total.
Hence, as in the $E^1$ case, we observe
the dominance of the pionic contribution.
Figure \ref{fi:delalpha} shows the
radial density of expression (\ref{eq:allampi}).
The long-range nature of the pion
is evident. As discussed in Sec. \ref{se:E1},
our predictions are enhanced due to
the fact that $g_A$ has a too large
value in our model. In the present case,
however, we estimate this effect at the level
of $g_A^{model}/g_A^{exp.} \sim 1.4$, rather than a
factor of $2$ in the pionic sea-gull term. This is because
Eq. (\ref{eq:allampi}) is linear in $\pi_h$
(recall that asymptotically $\pi_h$ is
proportional to $g_A$, Eq. (\ref{eq:pionasympt})),
and $\pi_S$ and $\pi_D$ are
not dependent on the strength of the pionic
tail in $\pi_h$. The reason is that the sources
in the linear response equations come
from the quarks, which are short-ranged.
In the asymptotic region, where
Eq. (\ref{eq:allampi}) gets most of
its contribution, the shifts in the pion
field, $\pi_S$ and $\pi_D$, depend on the
total strength of the source ($Q^{I=0} = \mbox{$1\over 2$}$),
and the pion mass, which enters the Green's
function, but not on the strength of the
tail of $\pi_h$. Consequently, our overestimate of $\delta\alpha$
should be at the level of 40\% only.

Our results are summarized in Table \ref{tab:alpha}.
We note, that the sign of the
neutron-proton splitting effect, and its
magnitude, agree with the recent
experimental numbers.

\subsection{Pionic cloud and sign of $\delta\alpha$}
\label{se:sign}

In the previous section the sign of
$\delta\alpha = \alpha_n - \alpha_p$ was
found to be negative. The following
plausibility argument can be given as to why
we expect this behavior in models with pionic clouds.
The hedgehog models imply that the
nucleon consists of a quark core, carrying isoscalar charge,
and a pion cloud, carrying isovector charge.
In the proton, these charges have the same sign, and the electric field
distorts both the core and the cloud, but it does not
displace them relative to each other. In
the case of the neutron, the cloud and the core, in addition to being
deformed also get displaced, since their charges are opposite.
This results in additional polarizability.
Calculations based on other methods
also give $\alpha_n > \alpha_p$ \cite{BKM91,Schroder}.

\subsection{Retardation term}
\label{se:retardation}

Our calculation was performed with a spatially constant
electric field. Now consider plane-wave photons, as in Compton
scattering, i.e.
\FL
\begin{eqnarray}
\mbox{\boldmath $E$} =
\mbox{\boldmath $E$}_0 exp(\dot{\imath}
\mbox{\boldmath $k$}_{\perp} \cdot \mbox{\boldmath $x$})
&\simeq& E_0 (1 + \dot{\imath} \mbox{\boldmath $k$}_{\perp}
\cdot \mbox{\boldmath $x$}- \mbox{$1\over 2$}
(\mbox{\boldmath $k$}_{\perp} \cdot \mbox{\boldmath $x$})^2),
\nonumber \\
{k_{\perp}}^2 &=& {\omega}^2.
\label{eq:retar}
\end{eqnarray}
The linear-response source, $j$,
is modified accordingly, and the charge
$Q = {\xi_0}^{\dagger} j$ (Sec. III in (II)) is replaced by an effective
charge
$Q^{eff.} = Q (1 - \mbox{$1\over 6$} {\omega}^2 {\langle r^2 \rangle}_E)$.
Substituting this expression to the
expression for the zero-mode part of the polarizability,
Eq. (\ref{eq:alphazero}), we obtain the usual retardation term,
exactly as in the classical derivation of ref. \cite{Ericson73}, as well
as in more general derivations \cite{Friar:rev,Petrunkin61}.

\section{Magnetic polarizability}
\label{se:magnetic}

Although naively one would think that
the analysis of the magnetic polarizabilities
would parallel the electric case, it
turns out that there are fundamental difficulties.
The reason is the non-commutativity of
the $\omega \rightarrow 0$ limit of the Compton
scattering, and the large-$N_c$ limit,
in the magnetic case. As stressed
earlier, since the large-$N_c$ limit is
the basic principle behind the hedgehog approach,
one has to work consistently in the $N_c$
counting. We point out that with the linear response
approach to the magnetic polarizabilities
it is not possible. We also show importance of pionic dispersive terms
in the magnetic polarizability, which enter
at the same level as the pionic sea-gull
contribution. Such terms were neglected in
previous works \cite{Nyman84,Chemtob87,Scoccola90}.

\subsection{The $N-\Delta$ Born term}
\label{se:ndelta}

The importance of the Born term with intermediate $\Delta$ state in
the estimates for $\beta$ is a well recognized fact. The hedgehog
possesses isovector magnetic moment, and the magnetic interaction term
in the effective hamiltonian has the form \cite{Nyman84,Scoccola90}
\begin{equation}
H^{magn.} = 3 \mu^{I=1} \mbox{\boldmath $c$} \cdot \mbox{\boldmath $B$},
\label{eq:magham}
\end{equation}
where $\mu^{I=1}$ is the isovector magnetic moment, and the collective
vector $\mbox{\boldmath $c$}$ is defined
in App. B in (II). Hamiltonian (\ref{eq:magham})
leads to $N - \Delta$ transitions, and, according to perturbation theory,
the contribution to the magnetic polarizability is
\begin{equation}
\beta_{N \Delta} = 18  (\mu^{I=1})^2
 \frac{| \langle N | c_0 | \Delta \rangle |^2}{M_{\Delta} - M_N} =
 \frac{4 (\mu^{I=1})^2}{M_{\Delta} - M_N},
\label{eq:betaND}
\end{equation}
where in the last equality we have used Eq. (II.B.9). Now, recalling the
$N_c$-counting rules \cite{ANW83,CB86}:
$\mu^{I=1} \sim N_c$, $M_{\Delta} - M_N \sim 1/N_c$, we immediately
obtain the result
\begin{equation}
\beta_{N \Delta} \sim N_c^3 .
\label{eq:betanc}
\end{equation}
This surprising ``superleading'' behavior is the result of collective
effects in the hedgehog wave function. In the orthodox approach to
$N_c$-counting, one should stop the analysis at this point, and conclude
that as far as the magnetic polarizability is concerned, one is not
close to the $N_c \rightarrow \infty$ limit: according to
Eq. (\ref{eq:betanc}) $\beta$ should be much larger than $\alpha$, which
contradicts the experiment. Using the physical values for various magnitudes in
Eq. (\ref{eq:betaND}) we obtain the numerical value
\begin{equation}
\beta_{N \Delta} \sim 12 \times 10^{-4} fm^3.
\label{eq:bndnum}
\end{equation}

Note, that to compare fairly to experiment, a given model
has to predict properly the value of $\mu_{N \Delta}$.

A related fundamental point is that
the polarizabilities are defined in $\omega \rightarrow 0$
limits in the Compton amplitude, and
in the physical world the resonances are at
finite values of $\omega$, well separated from the $\omega = 0$ region.
In the large-$N_c$ limit, the $\Delta$-resonance occurs
at $\omega = 0$. Thus, the meaning
of the polarizability is in fact lost, unless the
$\omega \rightarrow 0$ limit is taken
before the $N_c \rightarrow \infty$ limit. In
the linear response method, the limits are
implicitly taken in the reversed order.

In principle, a consistent analysis of the $N_c$-subleading physics is
possible, but this would require
a fully quantum-mechanical (not semiclassical)
treatment, e.g. one could use the Kerman-Klein method \cite{KermanKlein}.
Technically, this would be a tremendous effort, and of questionable merit
in a simple effective theory such as hedgehog models.

In order to be able to make some estimates,
we relax the strict $N_c$-counting requirement in the rest
of this section, and try to examine the subleading terms in $N_c$
in a more ``flexible'' approach.

\subsection{Sea-gull term}
\label{se:magnsg}

In hedgehog models, the magnetic seagull term is
related to the electric sea-gull term by
the relation (\ref{eq:polarsg}). In enters
at the level $N_c^1$. Numerically,
\begin{equation}
\beta_{sea-gull} = -14 \times 10^{-4} fm^3,
\label{eq:bsgnum}
\end{equation}
which would almost exactly cancel the
$\beta_{N \Delta}$ term, in apparent rough
agreement with experiment. This is, however, not
the full story at the level $N_c$. In the
next section we show, that the dispersive
contributions to $\beta$ arise of the same
order as $\beta_{sea-gull}$, but have positive sign,
and the desired cancellation is largely suppressed.

\subsection{Dispersive terms}
\label{se:magndisp}

The calculation of the magnetic dispersive
terms goes along the same lines as the
calculation of the electric polarizability.
There are, however, a few
important differences. Firstly, for the
ever-${\cal R}$ $B^1$ perturbation the
pionic term in Eq. (\ref{eq:lB11}) leads to
pionic sources in the linear response
equations. Since these are long-ranged
(proportional to $\pi_h$), we get strong
dispersive effects. The sources are listed
explicitly in Table \ref{tab:kpr}. For
the $K=0^{++}$ and $2^{++}$ cases we obtain
equations of the form (II.3.8). The only
difference is that in the $K=0^{++}$
the valence quark eigenvalue, $\varepsilon$,
changes, and the resulting quark
equation in (II.3.8) has the form
\begin{equation}
(h - \varepsilon) \delta q^{+} -
\delta\varepsilon q_h - 2 g M q_h Z = j^{+}_q ,
\label{eq:epsshift}
\end{equation}
where $\delta\varepsilon$ is the shift in the quark eigenvalue.
This shift may be viewed as a Lagrange
multiplier ensuring the orthogonality
of the shifted quark spinor,
$\delta q^{+}$, and the hedgehog quark spinor, $q_h$.
The numerical methods of treating the
$K=0^{++}$ and $2^{++}$ cases are straightforward.

In the $K=1^{++}$ case, however, the perturbation
excites the rotational zero-mode.
This is a similar mechanism as in the case of
the excitation of the translational
zero mode in the $E^0$ case ---
$K^{\cal PR} = 1^{++}$ are the quantum
numbers of the rotational mode in spin,
or isospin (these are equivalent, since
the hedgehog soliton is a $K^{\cal PR} = 0^{++}$
object). Thus, the hedgehog baryon
placed in a constant magnetic field starts
to rotate, and continues to spin up
indefinitely. Quantum-mechanically,
if time-dependent perturbation theory
were used, it would correspond to transitions to
higher spin and isospin states.

There are a few subtle issues here, concerning the
order of limits. Suppose we perform (exact) projection
first (e.g. suppose we have done a calculation
using the Kerman-Klein method \cite{KermanKlein}),
and have states with good angular momentum and
isospin on which we apply the magnetic
perturbation. Then, the rotational zero
modes no longer appear, since they arise only if
the solution breaks the symmetry of
the lagrangian (Sec. III B in (II)).
Thus, the appearance of the rotational zero
mode in the $K^{\cal PR} = 1^{++}$ is the
effect of starting from the unprojected
hedgehog. In fact, this problem is
another manifestation of the noncommutativity
of the $\omega \rightarrow 0$ and the
$N_c \rightarrow \infty$ limits, described in
Sec. (\ref{se:ndelta}). One may hope, that
in an improved treatment, the zero-mode
physics is described by the $N - \Delta$
term, Eq. (\ref{eq:betaND}). With
this in mind, in order to estimate the size
of the dispersive effects, we project out the divergent
zero-mode part from the $K=1$ contribution,
and retain the physical part. The total result
for $\beta$ consists of $\beta_{N \Delta}$,
$\beta_{sea-gull}$, and $\beta_{disp.}$ with
the zero-mode contribution projected out. As mentioned before, this
is not a consistent procedure, but it allows us to estimate the
importance of various effects in the magnetic polarizability.

The numerical values obtained for the
$K=0, 1$ and $2$ parts of the $B^1$ magnetic
perturbation are $3.6$, $-0.6$ and
$3.3 \times 10^{-4} fm^3$, respectively.
The total dispersive contribution to $\beta_N$ is
\begin{equation}
\beta_{disp} = 6.5 \times 10^{-4} fm^3,
\label{eq:bsgdisnum}
\end{equation}
The quark
contribution is negative, and carries $- 15 \%$ of the total.

The isoscalar magnetic perturbation,
$B^0$, which is a $1/N_c$ effect, and comes
entirely from the quarks, contributes $0.2 \times 10^{-4} fm^3$,
which is negligible, as expected
from $N_c$ counting. We also expect the neutron-proton
splitting of magnetic polarizabilities, $\delta\beta$,  to
be small. It comes entirely from the
quarks, since there is no magnetic analog of
the electric term (\ref{eq:llamE1}), and is expected
to be much smaller than $\delta\alpha$,
which had a pionic contribution.

At this point it might seem that adding up contributions
(\ref{eq:bsgnum},\ref{eq:bsgdisnum},\ref{eq:bndnum}) we still get
a partial cancellation. However, the sea-gull and dispersive pieces
are reduced if the pion field has the correct strength by a factor of
$\sim 2$, as in the electric sea-gull case.
The effects discussed in Sec. \ref{se:chipt}
further reduce these values, by another factor of 2.
These effects cause the cancellation to disappear, and we are left
with a large value of $\beta_N$. As stressed
before, rigorous estimates could
only be done in a calculation beyond linear response.

\section{Role of the $\Delta$ in hedgehog
models and chiral perturbation theory}
\label{se:chipt}

In this section we compare the hedgehog
model predictions to the results obtained
in chiral perturbation theory $\chi PT$.
At first, it may seem awkward, since
the two methods are based on two
different limits: $N_c \rightarrow \infty$,
and $m_\pi \rightarrow 0$, which
are known not to commute, and give
different results for various observables
\cite{Adkins84}. However, as
we show in Ref. \cite{CB92chirnt},
for quantities which are divergent in the chiral
limit as $m_\pi^{-1}$, there is a simple
connection between the hedgehog predictions and
the $\chi PT$ predictions for
scalar-isoscalar and vector-isovector quantities
(electromagnetic polarizabilities $\alpha_N$
and $\beta_N$ are scalar-isoscalar).

Let us evaluate the polarizabilities
using our methods in the chiral limit.
For $\alpha_N$, the dominant part comes
from the sea-gull term. In the chiral
limit the quantity diverges as $m_\pi^{-1}$,
and all of the contribution
comes from the pionic tail, which has the form
(\ref{eq:pionasympt}). Evaluating integral
(\ref{eq:polarsg}) with the profile (\ref{eq:pionasympt} we obtain
\begin{equation}
\alpha_N = \frac{5 e^2 g_A^2}{32 \pi M^2 m_{\pi}} .
\label{eq:alchi}
\end{equation}
Similarly, including both the sea-gull
and dispersive pieces for the magnetic
polarizability, we obtain in the chiral limit
\begin{equation}
\beta_N = \frac{e^2 g_A^2}{64 \pi M^2 m_{\pi}} .
\label{eq:bechi}
\end{equation}
These expressions are {\em exactly}
a factor of 3 larger than the
analogous expressions following from
$\chi PT$ \cite{BKM91}. This difference
comes from the different treatment
of the $\Delta$ isobar. In $\chi PT$ the
$\Delta$ contributions in pionic loops
(Fig. \ref{fi:oneloop}) are not
included, since it is implicitly
assumed that the $N - \Delta$ mass splitting is
much larger than the pion mass, and
consequently loops with the $\Delta$
do not contribute to the leading singularity. They
are counted as effects of order $log(m_\pi), 1, ...$. In
hedgehog models, on the contrary, the $N - \Delta$
mass splitting is a $N_c^{-1}$ effect,
much smaller than the pion mass, which is
of the order $N_c^{0}$. Therefore,
hedgehog models include the $\Delta$ on
equal footing with the nucleon. As
shown below, spin and isospin
Clebsch factors account for the
difference by a factor of 3
between Eqs. (\ref{eq:alchi},\ref{eq:bechi}),
and $\chi PT$ predictions.

One may ask, how the physics of pionic loops,
such as in Fig. \ref{fi:oneloop},
is present in hedgehog models. After all,
the treatment of the pion field
in hedgehog models is classical. To demonstrate
how hedgehog expressions
in the chiral limit may be viewed as hadronic
loop diagrams such as in Fig. \ref{fi:oneloop},
let us evaluate this diagram in the chiral limit.
First, we perform a calculation
of the self-energy due to an insertion of the
electric sea-gull interaction
in the pion loop. In momentum space, we obtain
\FL
\begin{eqnarray}
\Sigma^{*} &=& (\frac{g_A}{2 F_{\pi}})^2
\int \frac{d^{4} k}{(2 \pi)^4}
 \int \frac{d^{4} q}{(2 \pi)^4} \;
\tau_a \gamma_5 (\slash{k} + \mbox{$1\over 2$}
\slash{q}) S_{F}(p-k) \nonumber \\
& & \tau_b \gamma_5 (\slash{k} -
\mbox{$1\over 2$} \slash{q})
D(k - \mbox{$1\over 2$} q) \tilde{V}(q)
D(k + \mbox{$1\over 2$} q) T_{ab},
\label{eq:Feynman}
\end{eqnarray}
where  $S_{F}$ is the Feynman propagator of the nucleon,
$D(k) = \dot{\imath} /(k^2 - m_{\pi}^2 +
\dot{\imath} \varepsilon)$ is the propagator of the pion
field,
$\tilde{V}(q) =
\mbox{$1\over 2$} e^2  {(2 \pi)^4} \delta^4(q)
\mbox{$1\over 2$} (\mbox{\boldmath $E$} \cdot
\nabla_{\mbox{\boldmath $q$}})^2$
is the Fourier transform of the electric
sea-gull interaction in the coordinate space,
$V(\mbox{\boldmath $r$}) = \mbox{$1\over 2$}
e^2 (\mbox{\boldmath $E$} \cdot \mbox{\boldmath $r$})^2$, and
$T_{ab} = \delta_{ab} - \delta_{a3} \delta_{b3}$.
Now we proceed as follows:
we first carry out the integral over $k^0$ in
(\ref{eq:Feynman}). For the leading
chiral singularity piece, $m_{\pi}^{-1}$, the
contribution comes from the poles
in the pionic propagators. The poles in
the nucleon propagators contribute
to less singular terms. We can then perform
the nonrelativistic reduction
of the nucleon propagator and the pion-nucleon
vertices, and take the expectation
value of $\Sigma^{*}$ in positive energy spinors,
in order to extract the energy shift.
Transforming back to coordinate space, we obtain
the expression
\begin{equation}
\alpha =  e^2 \int d^{3} x \,
\phi_a^{asym.}(\mbox{\boldmath $r$})
\phi_b^{asym.}(\mbox{\boldmath $r$}) T_{ab} ,
\label{eq:sg5}
\end{equation}
which has the form of the expression obtained in hedgehog models.

We now go back to the question of the
factor of 3. When collective coordinates are
introduced, expression (\ref{eq:sg5}) becomes
\begin{equation}
\alpha =  e^2 \int d^3 x \,
\left ( \mbox{\boldmath $\phi$}^{asym.}
\times \mbox{\boldmath $c$} \right)^2 .
\label{eq:sg6}
\end{equation}
In Sec. \ref{se:seagull}, to obtain the
sea-gull contribution to the $\alpha_N$
we evaluated the collective matrix
of expression (\ref{eq:sg6}) in nucleon
collective wave functions.
Now we repeat the procedure, but we write
the resulting matrix element of the integrand in
Eq. (\ref{eq:sg6}) as
\begin{equation}
\sum_{i} \sum_{a} \langle N |(\mbox{\boldmath
$\phi$}^{asym.} \times \mbox{\boldmath $c$})_a | i \rangle
 \langle i |(\mbox{\boldmath $\phi$}^{asym.}
\times \mbox{\boldmath $c$} )_a | N \rangle ,
\label{eq:sg7}
\end{equation}
where $|i \rangle$ is a collective nucleon or $\Delta$
intermediate state. If the sum over $i$ is unrestricted, then
we just recover our previous expressions,
(\ref{eq:alchi},\ref{eq:bechi}).
If, however, we restrict $i$
to run only over nucleon collective states,
then using Eq. (B10) in (II), we
obtain a result smaller by a factor of 3,
and this result agrees with the
$\chi PT$ prediction. Conversely, had the
$\chi PT$ calculation included
the $\Delta$ in diagrams of Fig. \ref{fi:oneloop},
with $M_\Delta = M_N$, it
would predict a result 3 times larger than
quoted in Ref. \cite{BKM91}.

Let us introduce
\begin{equation}
d = (M_{\Delta} - M_N) / m_{\pi} .
\label{eq:d}
\end{equation}
These two cases, hedgehog models and $\chi PT$, correspond to
two limits: $d \rightarrow 0$, and $d \rightarrow \infty$.
In nature, $d \simeq 2$, which is between the two limits,
and we do not have the separation of the
pion mass and $M_{\Delta} - M_N$ scales.
In this case it seems most appropriate to treat both scales as small, and
keep $d$ as an unconstrained parameter.
This is the spirit of the approach
of Refs. \cite{JM,JM2}.
To estimate the contribution of the $\Delta$
at the physical value of $d$, we evaluate
the diagram of Fig. \ref{fi:oneloop},
starting from expression (\ref{eq:Feynman}),
with $S_F$ describing the $\Delta$ propagator
with the physical mass, and with the vertices
modified appropriately.
The $m_{\pi}^{-1}$ contribution is easily
obtained, since, as in the nucleon calculation
presented above, we can perform the non-relativistic
reduction. The result for the ratio
of the $\Delta$ to nucleon contribution in diagram \ref{fi:oneloop} is
\begin{equation}
\frac{\alpha^{\Delta}}{\alpha^N} = 2 S(d) ,
\label{eq:ratio}
\end{equation}
where the 2 comes from Clebsch factors, and the
``mass suppression'' function has the simple form
\begin{equation}
S(d) = \frac{4}{\pi} \left\{ \begin{array}{ll}
   {\rm Arctan}  \left( \sqrt{ \frac{1-d}{1+d} }
\right) / \sqrt{1-d^2} & \;\;{\rm for} \; d \leq 1 \\
   {\rm Arctanh} \left( \sqrt{ \frac{d-1}{1+d} }
\right) / \sqrt{d^2-1} & \;\;{\rm for} \; d > 1
   \end{array} \right.
\label{eq:masssupr}
\end{equation}
For degenerate $N$ and $\Delta$, $S(d=0)$ = 1. In the limit of
$d \rightarrow \infty$, $S(d) \sim \log{d}/d$.
For the physical value of $d$ we find $S(2.1) = 0.47$,
and from Eq. (\ref{eq:ratio}) we find, that the
$\Delta$ contribution in pionic loops of Fig. \ref{fi:oneloop}
to the electric polarizability is roughly equal to the nucleon
contribution. The results for the magnetic polarizability are analogous.

This important feature of large $\Delta$ contributions in pionic loops
(for scalar-isoscalar operators) is also present
in other quantities \cite{CB92chirnt}. In the
above discussion we have tacitly assumed that the ratio
$g_{\pi N \Delta} / g_{\pi NN} = 3/2$ \cite{Adkins84,CB86},
as predicted by hedgehog models. Experimental
numbers agree with this prediction to within
a few percent. The correction may be introduced to
expression (\ref{eq:ratio}) \cite{CB92chirnt}.

We note that the above analysis and the comparison
of the hedgehog model predictions and
the $\chi PT$ predictions can be done
only for observables which do not depend on cranking
(independent of the cranking frequency, $\lambda$).
Observables, which do depend on cranking
do not have the same chiral singularities,
e.g. the electric mean squared radius
diverges as $m_{\pi}^{-1}$ in hedgehog models,
and as $\log{m^{\pi}}$ in $\chi PT$. The same
is true of the splitting of electric polarizabilities, $\delta\alpha$.
In this case the issue of nonocommutativity of the
large-$N_c$ and chiral limits is much more complicated.

The results of this section indicate how one should try to
improve the hedgehog predictions by subtracting the amounts
by which these models overestimate the leading chiral singularity
terms. There is uncertainty in such a non-rigorous procedure, but
in our view it is required by physics. As already mentioned at the end
of Sec. \ref{se:ndelta}, the correct treatment of the $\Delta$
could be done in a quantum calculation
with the inclusion of the $N_c$-subleading terms.

On the other hand, the predictions of $\chi PT$ will be largely
effected by the diagrams of Fig. \ref{fi:oneloop} with $\Delta$
intermediate states. For polarizabilities, at the
leading singularity level, these
diagrams are as important as the diagrams
with the nucleon intermediate states.

\section{Effects of the pionic substructure}
\label{se:substructure}

The minimal substitution prescription in an effective
lagrangian cannot produce all interactions of a hadronic system
with the electromagnetic field.
It is possible to write down terms which
are by themselves gauge-invariant, and
thus not obtainable by gauging lagrangian (\ref{eq:GML}).
In this section we focus on
terms ${\cal L}_9$ and  ${\cal L}_{10}$ of reference \cite{Gasser}:
\FL
\begin{eqnarray}
{\cal L}_9 &=& -\dot{\imath} L_9 \,
Tr[{\cal F}_{\mu \nu}^L D^{\mu} U D^{\nu}U^{\dagger} +
      {\cal F}_{\mu \nu}^R D^{\mu}
U^{\dagger} D^{\nu} U ] , \\
{\cal L}_{10} &=& L_{10} \,
Tr[{\cal F}_{\mu \nu}^L U {\cal F}^{\mu \nu , R} U^{\dagger}] ,
\label{eq:Lag910}
\end{eqnarray}
where ${\cal F}_{\mu \nu}^{L,R}$ are
the left and right chiral field strength
tensors. For the case of electromagnetic field we have
${\cal F}_{\mu \nu}^{L,R} = e \mbox{$1\over 2$} \tau_3 {F}_{\mu \nu} =
e \mbox{$1\over 2$} \tau_3 (\partial_{\mu} A_{\nu} - \partial_{\nu} A_{\mu})$.
In the linear $\sigma$-model,
$U$ corresponds describes the
chiral field, $U = F_{\pi}^{-1} (\sigma +
\dot{\imath} \mbox{\boldmath $\tau$} \cdot
\mbox{\boldmath $\pi$})$. The constants $L_9$ and $L_{10}$ can be
expressed through measurable quantities,
namely the pion electric mean square
radius, ${\langle r^2 \rangle}_{E}^{\pi}$,
and the pion polarizability
$\overline{\alpha}^{\pi}= - \overline{\beta}^{\pi}$ \cite{Gasser,Holstein}:
\begin{eqnarray}
{L}_9 &=& \frac{F_{\pi}^2 {\langle r^2 \rangle}_{E}^{\pi}}{12} , \\
{L}_{10} &=& \frac{m_{\pi} F_{\pi}^2 \overline{\alpha}^{\pi}}{4} - L_9 =
\frac{m_{\pi} F_{\pi}^2 \alpha^{\pi}}{4} .
\label{eq:L910}
\end{eqnarray}
In the chiral perturbation theory treatment,
one considers pionic loops, and
through renormalization ${L}_9$ and  ${L}_{10}$
acquire chiral logarithms. In our
approach, we simply treat the terms
(\ref{eq:Lag910}) in the mean-field
approximation, replacing the meson
field operators by classical fields.
As before, the $E$ and $B$ fields are constant,
and we find from the ${\cal L}_9$
term
\FL
\begin{eqnarray}
\int d^3 x \,{\cal L}_9 &=& 4 e^2 L_9 F_{\pi}^{-2}
\int d^3 x \, F^{i \nu} A_{\nu}
  (\mbox{\boldmath $c$} \times \mbox{\boldmath $\pi$}_h) \cdot
 (\mbox{\boldmath $c$} \times \partial_i
\mbox{\boldmath $\pi$}_h) \nonumber \\
&=& 2 L_9 (E^2 - B^2) \int d^3 x
(\mbox{\boldmath $c$} \times \mbox{\boldmath $\pi$}_h)^2  ,
\label{eq:L9int}
\end{eqnarray}
where $i$ runs over spatial values. The ${\cal L}_{10}$ term leads to
\FL
\begin{eqnarray}
\int d^3 x \,{\cal L}_{10} &=& 4 e^2 L_{10}
(E^2 - B^2)  F_{\pi}^{-2} \nonumber \\
& &\int d^3 x \, \left (  (\mbox{\boldmath $c$}
\times \mbox{\boldmath $\pi$}_h)^2 -
\mbox{$1\over 2$} (\sigma_h^2 + \mbox{\boldmath $\pi$}_h^2) \right ) .
\label{eq:L10int}
\end{eqnarray}
The last term under the integrand is canceled by
the vacuum subtraction, which is implicit.
It becomes
$- (\sigma_h^2 + \mbox{\boldmath $\pi$}_h^2) +
\sigma_{vac}^2 = - (\sigma_h^2 + \mbox{\boldmath
$\pi$}_h^2) + F_{\pi}^2$,
which is zero in the nonlinear sigma model,
but also  vanishingly small in our case
due to proximity of the hedgehog solution to
the chiral circle. We can thus drop this
term in our estimate. Using (\ref{eq:L910}, II.3.21)  we read off from
(\ref{eq:L9int}, \ref{eq:L10int}) the following contribution to the nucleon
polarizabilities:
\begin{equation}
\alpha_N^{\pi} = -\beta_N^{\pi} = m_{\pi} \Theta^{mes} {\overline\alpha^{\pi}}
{}.
\label{eq:alphaNpi}
\end{equation}
The physical interpretation of this
contribution is clear. The pion, having
electromagnetic structure, is polarizable.
Since the nucleon is surrounded
by a pion cloud, this pion polarizability
results in additional polarizability of the
nucleon. Note the opposite signs of the electric
and magnetic polarizabilities in
(\ref{eq:alphaNpi}), reflecting the fact that
$\overline{\alpha^{\pi}} = - \overline{\beta^{\pi}}$. Also notice, that
since ${\overline\alpha^{\pi}}$ is of order
1 in $N_c$-counting, the contributions
\ref{eq:alphaNpi} are of order $N_c$. The terms (\ref{eq:Lag910}) do
not lead to additional neutron-proton splitting of polarizabilities

We can call $N_{\pi}=m_{\pi} \Theta^{mes}$
in \ref{eq:alphaNpi} the ``number of pions''
in the nucleon seen in the Compton
scattering process. Numerically, $N_{\pi} \simeq 0.5$ in our model.
Using the relation of the moment of inertia to the $N-\Delta$ mass
splitting, we can write
\begin{equation}
N_{\pi} = \frac{3}{2} \frac{m_{\pi}}{M_{\Delta} - M_N}
\frac{\Theta^{mes}}{\Theta},
\label{eq:Npi}
\end{equation}
where the last factor is the fraction of the total moment of
inertia carried by the pion. This quantity
is $\sim 1$ in hedgehog models:
60\% in our model with quarks, 100\% in the Skyrmion.
We thus have a quasi-model-independent
result $N_{\pi} \sim 0.5 - 0.7$.

The value of ${\alpha^{\pi}}$ can be determined
experimentally, however existing
experimental data \cite{pionpolexp} do not seem
reliable, and are in
contradiction with a low-energy theorem due to
Holstein \cite{Holstein}, which gives
${\alpha^{\pi}} = 2.8 \times 10^{-4} fm^3$. With this value
we get
\begin{equation}
\alpha_N^{\pi} = -\beta_N^{\pi} \sim 1.3 \times 10^{-4} fm^3 ,
\label{eq:alphaNpinum}
\end{equation}
which is a few times smaller compared to the minimal substitution terms, but
non-negligible, especially for the magnetic case, where we expect
cancellations to occur. If experimental numbers for $\alpha^{\pi}$ were used
\cite{pionpolexp}, then a three times larger result would follow.

Other non-minimal substitution terms can also be considered, but there is no
knowledge of the phenomenological low-energy
constants which have to be introduced,
and predictive power is lost.

\section{Other models}
\label{se:othermodels}

The analysis of this paper can be straightforwardly applied to
other chiral models. In purely mesonic
Skyrmions \cite{skyrmion:rev}, the RPA
approximation corresponds to linearizing the
small fluctuation equations, hence the non-linearity constraint
is not imposed at the quantum level. Appropriate linear
response equations can be derived along the lines of App. B in (II).
The role of the isoscalar source is carried
by the topological (Goldstone-Wilczek)
current, and effects of the higher order terms
in the equations are present.

In purely quark models (NJL) \cite{NJL}, upon minimal
substitution, the lagrangian
in presence of the electromagnetic
field $A^{\mu}$, and after introducing
collective degrees of freedom, becomes
\FL
\begin{eqnarray}
S_{NJL}(A^{\mu}) &=& - \dot{\imath} \,Tr
\,log \left ( \dot{\imath} \slash{\partial{}}
                   - g (\sigma + \dot{\imath}
\gamma_5 \mbox{\boldmath $\tau$} \cdot \mbox{\boldmath $\pi$})
                   - \mbox{$1\over 2$} \gamma_0
\mbox{\boldmath $\tau$} \cdot \mbox{\boldmath $\lambda$} \right.
\nonumber \\
                 &+& \left. \mbox{$1\over 2$} \slash{A} (N_c^{-1} +
\ \mbox{\boldmath $\tau$} \cdot \mbox{\boldmath $c$}) \right )
                   - (vac) ,
\label{eq:NJLem}
\end{eqnarray}
where $(vac)$ denotes the vacuum subtraction,
and an NJL cut-off is understood.
The trace accounts for the occupied (valence) levels. This
expression may be expanded to second-order in $A^{\mu}$, and
expressions for polarizabilities can
be easily derived. The dispersive pieces,
as in the model treated in this paper, lead in case of even-${\cal R}$
interactions to distortions in the profile
functions $\sigma$ and $\mbox{\boldmath $\pi$}$.
This is likely to result in numerical complications, in particular in
cases with zero-mode excitations. One can use gradient expansion
techniques \cite{grad} instead of solving the model exactly. Then,
for example, the isovector electric responce of the quarks from
the Dirac sea generates the pionic sea-gull contribution.

All our generic hedgehog model conclusions described in this paper
hold for these, and other hedgehog models.

\section{Conclusion}
\label{se:conclusion}

The main results obtained in this paper can be summarized as follows:

1) We show the $N_c$-counting rules for electromagnetic polarizabilities
in hedgehog models. For electric polarizabilities the basic experimental
pattern
is reproduced ($\alpha_N \sim N_c > \delta\alpha \sim 1/N_c$).
For the magnetic case the rules show the inapplicability
of linear response ($\beta_N \sim N_c^3$).

2) Dispersive terms lead to deformation of hedgehog solitons, and
the resulting contributions to polarizabilities enter at the same $N_c$-level
as the sea-gull contributions. In the magnetic case, we expect that
the presence of dispersive terms leads to large positive
contributions to $\beta_N$. There is no cancellation mechanism which can
bring $\beta$ down to the experimental value. In fact, all model
calculations with the $\Delta$ degree of freedom
have this problem, since the paramagnetic $N$-$\Delta$
term is large, and there is no simple mechanism to cancel
this effect.

3) Hedgehog models provide a mechanism of splitting  of the neutron and
proton polarizabilities. An explicit
calculation gives reasonable numbers, and the sign
of $\delta\alpha = \alpha_n - \alpha_p$ is expected to be positive
in models with pionic clouds.

4) Concerning numerical results in hedgehog models, because of the sensitivity
of the results to the pion tail, the value of $g_A$ in a model should be
well reproduced. Also, for the magnetic case, good prediction for
$\mu_{N \Delta}$, as well as $M_{\Delta} - M_N$,
is necessary to reproduce the $N - \Delta$
paramagnetic term.

5) The $\Delta$ resonance plays an important role. Hedgehog models
largely overestimate these contributions, since they neglect the
effects of $N - \Delta$ mass splitting in projection. We show
how to estimate the $\Delta$ effects in pionic loops in a modified
chiral perturbation theory, and find these effects are at the level of 100 \%
in calculation of the electromagnetic polarizabilities. One should also try
to improve the hedgehog predictions by subtracting the amounts
by which these models overestimate the leading chiral singularity
terms.

6) The effects of non-minimal substitution terms ${\cal L}_9$ and
${\cal L}_{10}$ \cite{Gasser} in the effective lagrangian
enter at the level of $1 - 2 \times 10^{-4} fm^3$.

\acknowledgements

Support of the the National Science Foundation (Presidential Young
Investigator grant), and of the U.S. Department of Energy is gratefully
acknowledged. We thank Manoj Banerjee for many useful
suggestions and countless valuable comments.
One of us (WB) acknowledges a partial support of
the Polish State Committee for Scientific Research
 (grants 2.0204.91.01 and 2.0091.91.01).

\newpage


\newpage
\widetext

\figure{Dispersive contributions to the
 Compton scattering amplitude. In our treatment,
${\cal G}$ is the RPA (linear
 response) propagator.
 \label{fi:rpa}}

\figure{Radial densities of the leading-$N_c$
 contributions to $\alpha_N$ (in units of $10^{-4} fm^{2}$):
 Sea-gull (solid line), and quark $E^1$ contribution (dashed line).
 The dotted line shows the sea-gull
term evaluated with the asymptotic
 profile, Eq. (\ref{eq:pionasympt}).
   \label{fi:radsg}}

\figure{Radial density of the pion contribution
to the splitting of the neutron and proton
  electric polarizabilities, $\delta\alpha$ (in units of $10^{-4} fm^{2}$).
   \label{fi:delalpha}}

\figure{Hadronic one-pion-loop diagrams
giving the leading chiral contribution to
 nucleon polarizabilities. The vertex
corresponds to the sea-gull interaction
 $e^2 \, \int d^3 x \, (\mbox{\boldmath $E$} \cdot \mbox{\boldmath $x$})
 (\mbox{\boldmath $c$} \times \mbox{\boldmath $\pi$})$
   \label{fi:oneloop}}

\newpage
\widetext

\widetext

\begin{table}
\caption{$K^{PR}$ classification of
various dispersive perturbations, and
sources in the corresponding linear
response equations (App. A in (II)). Quantum
numbers $(L, \Lambda)$ label quark
shifts, $L_{\pi}$ and $L_{\sigma}$ are
the orbital numbers of meson shifts.
Null entry in columns for
$L_{\pi}$ and $L_{\sigma}$ means that
the fluctuation does not arise.
For cases with zero modes (electric
isoscalar and magnetic isovector,
{K=1}), the sources listed are twice
the $j_X$, $j_Y$ or $j_Z$ sources from
equations in App. A in (II). For other
cases they are the sources entering
Eqs. (II.3.7,II.3.8). Radial functions $G_h$ and $F_h$
are the upper and lower components
of the hedgehog valece spinor, and $\pi_h$
in the hedgehog pion profile. See (II) for details.}
\begin{tabular}{ll|lcc|lc|lc}
 &  & \multicolumn{3}{c|}{Quarks} & \multicolumn{2}{c|}
{$\delta \pi$} & \multicolumn{2}{c}{$\delta \sigma$}  \\
interaction &  $K^{\cal PR}$ & & & & & &  \\
 & &(L,$\Lambda$) & $j_G/ \sqrt{4 \pi}$ & $j_F/ \sqrt{4 \pi}$
& $L_{\pi}$ & $j_{\pi}/ \sqrt{4 \pi}$ & $L_{\sigma}$
& $j_{\sigma}/ \sqrt{4 \pi}$  \\
\hline
\hline
& & & & & & & &  \\
cranking           & $1^{+-}$ & (0,1)     &
$-\mbox{$1\over 2$} G_h$                 &
$-\mbox{$1\over 2$} F_h$      & &   & &     \\
                   &          & (2,1)     &
                $0$              &
      $0$        & &   & &     \\
& & & & & & & &  \\
\hline
& & & & & & & &  \\
electric isovector & $0^{--}$ & (1,1)
   & $\mbox{$1\over 2$} r G_h / \sqrt{3}$
    & $\mbox{$1\over 2$} r F_h / \sqrt{3}$ & &   & &     \\
& & & & & & & &  \\
                   & $1^{--}$ & (1,0)     &
                 $0$              &
           $0$     & &   & &     \\
                   &          & (1,1)
& $\mbox{$1\over 2$} r G_h / \sqrt{3}$
    & $\mbox{$1\over 2$} r F_h / \sqrt{3}$ & &   & &     \\
& & & & & & & &  \\
                   & $2^{--}$ & (1,1)
  & $\mbox{$1\over 2$} r G_h / \sqrt{3}$
   & $\mbox{$1\over 2$} r F_h / \sqrt{3}$ & &   & &     \\
                   &          & (3,1)
 &                    $0$             &
                  $0$     & &   & &     \\
& & & & & & & &  \\
\hline
& & & & & & & &  \\
electric isoscalar & $1^{-+}$ & (1,0)
  &                    $0$
&                            $0$
 & $0$ & $0$ & $1$ & $0$ \\
                   &          & (1,1)
   & $\mbox{$1\over 2$} N_c^{-1} r G_h / \sqrt{3}$ &
 $\mbox{$1\over 2$} N_c^{-1} r F_h / \sqrt{3}$
& $2$ & $0$ & &  \\
& & & & & & & &  \\
\hline
& & & & & & & &  \\
magn. isovector & $0^{++}$ & (0,0) &
 $\mbox{$1\over 2$} r F_h / \sqrt{3}$
& $\mbox{$1\over 2$} r G_h / \sqrt{3}$
& $1$ & $-2 \pi_h / \sqrt{3}$ & $0$
 & $0$  \\
& & & & & & & &  \\

                   & $1^{++}$ & (0,1)
  & $ rF_h / \sqrt{18}$ & $r G_h / \sqrt{18}$
& $1$ & $ - \pi_h / \sqrt{3}$  &  &  \\
                   &          & (2,1)
 & $-(1/12) r F_h$ & $-(1/12) r G_h$
 & &              &  &                \\
& & & & & & & &  \\

                   & $2^{++}$ & (2,0)
 & $\mbox{$1\over 2$} r F_h / \sqrt{30}$
& $\mbox{$1\over 2$} r G_h / \sqrt{30}$ &
 $1$ & $ \pi_h / \sqrt{3}$ & $2$ & $0$  \\
                   &          & (2,1)
 & $\mbox{$1\over 2$} r F_h / \sqrt{20}$ &
$\mbox{$1\over 2$} r G_h / \sqrt{20}$ & $3$ & $0$ &
&               \\
& & & & & & & &  \\
\hline
& & & & & & & &  \\
magn. isoscalar & $1^{+-}$ & (0,1)
 & $-(1/6) N_c^{-1} r F_h$ & $-(1/6) N_c^{-1} r G_h$
& &           & &            \\
                   &          & (2,1)
& $-(\sqrt{2}/12) N_c^{-1} r F_h$
& $-(\sqrt{2}/12) N_c^{-1} r G_h$  & &   &     \\
& & & & & & & &
\end{tabular}
\label{tab:kpr}
\end{table}

\narrowtext

\begin{table}
\caption{Electric polarizability. The model predictions for
$\alpha_N$ are expected to be largly reduced by
effects disussed in the text.}
\begin{tabular}{ll}
\multicolumn{2}{c}{$\alpha_N$ ($10^{-4} fm^3$)}     \\
\hline
sea-gull          &  28.5            \\
dispersive $E^1$ (quarks) &  3.5     \\
total             &  32             \\
experiment        &  9.6 $\pm$ 1.8 $\pm$ 2.2         \\
\hline
\multicolumn{2}{c}{$\delta \alpha$ ($10^{-4} fm^3$)}\\
\hline
pion             & 5.6              \\
quarks           & -0.2              \\
total            & 5.4              \\
experiment       & 4.8 $\pm$ 1.8 $\pm$ 2.2    \\
\end{tabular}
\label{tab:alpha}
\end{table}

\newpage


\begin{references}
\bibitem{remark:neutral}
{Strictly speaking, this definition
is for charge-neutral particles only, but
  can be easily modified for
the case of charged objects (see discussion in
  Sec. \ref{se:E0}}.
\end{references}
\end{document}